\documentclass[prd,aps,showpacs, tightenlines, 
nofootinbib,preprint,preprintnumbers
               ]{revtex4}

\usepackage{graphicx}
\usepackage{amsfonts}
\usepackage{amssymb}
\usepackage{latexsym}
\usepackage{cancel}
\usepackage{color}
\newcommand{\beq}{\begin{equation}}
\newcommand{\beqa}{\begin{eqnarray}}
\newcommand{\eeq}{\end{equation}}
\newcommand{\eeqa}{\end{eqnarray}}

\newcommand{\simg}{\gtrsim}
\newcommand{\siml}{\lesssim}
\newcommand{\lkk}{\left[}  \newcommand{\rkk}{\right]}

\def\JL#1#2#3{JETP. Lett. {\bf #1}, #2 (19#3)}

\def\MNRAS#1#2#3{Mon. Not. R. Astron. Soc. {\bf #1}, #2 (19#3)}

\def\PRDD#1#2#3{Phys. Rev. D {\bf #1}, #2 (20#3)}

\def\RMP#1#2#3{Rev. Mod. Phys. {\bf #1}, #2 (19#3)}

\begin{document}

\title{A Consistency Relation in Cosmology
}

\author{Takeshi Chiba}%
\affiliation{
Department of Physics, College of Humanities and Sciences, \\
Nihon University, 
Tokyo 156-8550, Japan}
\author{Ryuichi Takahashi\footnote{Address after April 1: 
Department of Physics and Astrophysics, Nagoya University, 
Chikusa, Nagoya 464-8602, Japan}
}
\affiliation{
Division of Theoretical Astronomy, National
 Astronomical Observatory of Japan, \\
Mitaka, Tokyo 181-8588, Japan}

\date{\today}
\preprint{astro-ph/0703347}

\begin{abstract}
We provide a consistency relation between cosmological 
observables in general relativity without relying on the 
equation of state of dark energy. 
The consistency relation should be satisfied if general relativity is  
the correct theory of gravity and dark energy clustering is negligible. 
As an extension, we also provide the DGP counterpart of the relation.
\end{abstract}


\maketitle

\section{Introduction}

General relativity has passed the experimental tests on the scales of 
the solar system with flying colors. 
Probing general relativity on cosmological scales \cite{jim} will be the next 
target for gravitational physics. In this paper, we take a small step toward this aim. 
We understand that such a program suffers from fundamental degeneracy 
between gravity theories and properties of dark energy\cite{kunz}. 
Rather than emphasizing the degeneracy, however, we focus on a consistency 
relation which tests the conventional theoretical framework in cosmology: 
general relativistic CDM model with (almost homogeneous) dark energy.  
Breaking the relation would be the signature of the breakdown of general relativity 
at the cosmological scale or non-negligible dark energy clustering. 

A consistency test has recently been proposed by Knox et al.\cite{knox} and 
Ishak et al.\cite{spergel} (see also \cite{uzan}). 
They look at a consistency between the expansion rate 
determined by the distance-redshift relation and that by the growth rate of 
large scale structure. An inconsistency would arise within the dark energy 
parameter space if the underling gravity theory is different from general relativity. 
Our work is partly motivated by these study, and we shall provide an explicit 
consistency relation  (without referring to the equation of state of dark energy) 
in general relativity which relates the 
two cosmological observables: the (luminosity) distance and 
the density perturbation. The basic idea is very simple. 
The expansion rate is determined both 
from the distance-redshift relation \cite{nc,alex} and from density perturbations \cite{alex}. 
Equating them then gives a consistency relation. 
It should hold if general relativity is the correct theory of gravity 
in the universe and if dark energy clustering is negligible. 
As an extension of the program we also provide a consistency relation 
in the DGP model.

\section{Reconstructing the Expansion Rate of the Universe}

\subsection{From Standard Candles}

The observations of type Ia supernovae, for example, yield the luminosity distance $d_L(z)$ 
through $m-M=5\log_{10}(d_L(z)/{\rm Mpc})+25$, with $m$ being the apparent magnitude 
and $M$ being the absolute magnitude. 
The luminosity distance is then related to the Hubble parameter $H(z)$ 
kinematically (assuming the energy conservation of photons) as 
\beqa
d_L(z)={(1+z)\over H_0\sqrt{|\Omega_{K0}|}}\sin_K\left(H_0\sqrt{|\Omega_{K0}|}
\int^z_0{dz\over H(z)}\right),
\eeqa
where $\Omega_{K0}=-K/a_0^2H_0^2$ and $\sin_K(x)=\sin(x)~ (K=1),x~(K=0),\sinh(x)~(K=-1)$. 
In terms of $r(z)=d_L(z)/(1+z)$,  the Hubble parameter is rewritten as \cite{nc}
\beqa
\left({H(z)\over H_0}\right)^2={1+r(z)^2H_0^2\Omega_{K0}\over H_0^2(dr(z)/dz)^2}.
\label{hubble:dl}
\eeqa
This is the first expression of $H(z)$ in terms of observables. Since it is purely kinematical 
relation, it holds in any metric theories of gravity (again assuming photon energy conservation). 

\subsection{From Density Perturbation: Consistency Relation in General Relativity}

The measurements of weak gravitational lensing (cosmic shear) give the information of linear 
density perturbations  (or linear growth rate). 


In general relativity, a linear density  
perturbation $\delta(z)$ at scales much smaller than the hubble radius 
obeys the following differential equation 
\beqa
\ddot \delta+2H\dot\delta-4\pi G\rho_M\delta=0,
\label{linear:gr}
\eeqa
where a dot denotes the derivative with respect to the cosmic time and
 $\rho_M$ is the matter density. 
Here we have assumed the perturbative properties 
of dark energy to some extent: dark energy does not have unusual sound speed and 
has negligible anisotropic stress and has negligible interaction with dark matter.
\footnote{If we allow either of them, the program will fail: 
both general relativity and DGP can have the same expansion rate and growth rate \cite{kunz}.}
According to Starobinsky \cite{alex}, we rewrite Eq.(\ref{linear:gr}) 
by changing the argument from the cosmic time to the scale factor $a$: 
\beqa
{dH(a)^2\over da}+2\left({3\over a}+{d^2\delta/da^2\over d\delta/da}\right)H(a)^2
={3\Omega_{M0}H_0^2a_0^3\delta\over a^5 (d\delta/da)},
\eeqa
where $\Omega_{M0}$ is the matter density parameter of today. 
By regarding the above equation as the differential equation for 
$H(a)^2$, the solution is obtained as \cite{alex}
\beqa
\left({H\over H_0}\right)^2={3\Omega_{M0}a_0^3\over a^6(d\delta/da)^2}
\int^a_0a\delta{d\delta\over da}da
=3\Omega_{M0}{(1+z)^2\over \delta'(z)^2}\int_z^{\infty}{\delta\over 1+z}
(-\delta')dz,
\label{hubble:linear}
\eeqa 
where a prime denotes the derivative with respect to $z$. In the above solution, 
a homogeneous is disregarded by taking only growing solution. 
Putting $z=0$ in the solution, we obtain 
\beqa
1={3\Omega_{M0}\over \delta'(0)^2}\int^{\infty}_0{\delta\over 1+z}
(-\delta')dz,
\label{omega}
\eeqa
which expresses $\Omega_{M0}$ in terms of $\delta$. $\delta(z)$ at higher redshift ($z\simg 5$) 
may not be determined from observations. However, in practice $\delta(z)$ can be well 
approximated as $\delta\propto 1/(1+z)$ there. 
Using Eq.(\ref{omega}), Eq.(\ref{hubble:linear}) can then be alternatively written as 
\beqa
\left({H\over H_0}\right)^2=
{(1+z)^2\delta'(0)^2\over \delta'(z)^2}
\lkk
1-{\int^z_{0}{\delta\over 1+z}
(-\delta')dz\over \int^{\infty}_{0}{\delta\over 1+z}
(-\delta')dz}
\rkk    ,
\label{hubble:linear2}
\eeqa
Equating Eq.(\ref{hubble:linear2}) with Eq.(\ref{hubble:dl}) gives the 
consistency relation between observables: 
\beqa
{1+r(z)^2H_0^2\Omega_{K0}\over H_0^2(dr(z)/dz)^2}=
{(1+z)^2\delta'(0)^2\over \delta'(z)^2}
\lkk
1-{\int^z_{0}{\delta\over 1+z}
(-\delta')dz\over \int^{\infty}_{0}{\delta\over 1+z}
(-\delta')dz}
\rkk      ,
\label{consistency}
\eeqa
which should hold if general relativity is the correct theory of gravity 
in the universe. It should be noted that in deriving Eq.(\ref{consistency}), 
we do not assume the Friedmann equation and hence we do not specify the equation of state 
of dark energy but specify the perturbative properties of dark energy (sound speed and 
anisotropic stress). Eq.(\ref{consistency}) thus tests the underlying 
gravitational theory modulo these assumptions of dark energy.\footnote{Linder introduced  
the gravitational growth index $\gamma$ defined by $d\ln (\delta/a)/d\ln a=\Omega_M(a)^{\gamma}-1$ 
and found that it is insensitive to the equation of state of dark energy \cite{linder}.}

The degeneracy with the curvature parameter $K$ may be broken 
by using the CMB shift parameter \cite{be}: 
\beq
R=\sqrt{{\Omega_{M0}\over |\Omega_{K0}|}}\sin_K
\left(H_0|\Omega_{K0}|\int_0^{z_{LSS}}{dz\over H(z)}\right),
\label{shift}
\eeq
where $z_{LSS}=1089$, 
which is measured to be $R=1.70\pm 0.03$\cite{wm}.

\subsection{Modified Gravity: DGP}

If the consistency relation Eq.(\ref{consistency}) would not hold, then  
dark energy has anisotropic pressure or interaction with dark matter \cite{kunz}, or 
general relativity is not the correct theory of gravity on cosmological scales. 
In this section, we focus on the latter possibility and look for another relation in 
gravitational theories other than general relativity. 
While Eq.(\ref{hubble:dl}) 
holds in any theories of gravity as mentioned earlier,  the modification of gravity theories 
affects the gravitational instability and hence modifies the third term 
in Eq.(\ref{linear:gr}): the self-gravity of density perturbations. 

For example, in scalar-tensor theories of gravity, the growth of density perturbations 
is modified simply as \cite{ncs}
\beq
\ddot\delta+2H\dot\delta-4\pi G_{\rm eff}\rho_M\delta=0,
\label{linear:st}
\eeq
where $G_{\rm eff}$ is the effective local gravitational "constant" measured by Cavendish-type 
experiment and is time dependent. 
The modified evolution equation of density perturbations in general may be written as
\beqa
\ddot\delta+2H\dot\delta-4\pi G\rho_M\left(1+{1\over 3\beta}\right)\delta=0,
\label{linear:modified}
\eeqa
where $\beta$ in general depends on time and is determined once we specify 
the modified gravity theory. 

As a concrete example, we consider DGP(Dvali-Gabadadze-Poratti) model \cite{dgp}.
DGP model is a model of brane world 
in which three-dimensional brane is embedded in an infinite five-dimensional 
spacetime (bulk). The action for the five-dimensional theory is 
\beqa
S={1\over 2}M_{(5)}^3\int d^4xdy\sqrt{-g_{(5)}}R_{(5)}
 +{1\over 2}M_{(4)}^2\int d^4x \sqrt{-g_{(4)}}R_{(4)}+S_m,
\eeqa 
where the subscripts 4 and 5 denote the quantities on the brane and in the bulk, 
respectively, $M_{4}$($M_{(5)}$) is the four(five)-dimensional reduced Planck mass, 
and $S_m$ is the action for matter on the brane. 

In DGP model, the Friedmann equation is modified as \cite{def}
\beqa
H^2+{K\over a^2}={8\pi G\over 3}\left(\sqrt{\rho_M+\rho_{r_c}}+\sqrt{\rho_{r_c}}\right)^2,
\label{friedmann:dgp}
\eeqa
where $\rho_{r_c}=3/(32\pi Gr_c^2)$. $r_c$ is related to  
 the four dimensional Planck mass $M_{(4)}$ and the five dimensional
 Planck mass $M_{(5)}$ as $r_c=M_{(4)}^2/2M_{(5)}^3$. 
The evolution of density perturbations is also modified 
and $\beta$ in Eq.(\ref{linear:modified}) is given by \cite{koyama}:
\beqa
\beta=1-2r_cH\left(1+{\dot H\over 3H^2}\right).
\label{linear:dgp}
\eeqa
If we use the Friedmann equation Eq.(\ref{friedmann:dgp}), $\beta$ can be written as 
\footnote{The use of the Friedmann equation (theory) may be  out of the spirit of 
the program of the consistency relation: determine $H(a)$ only from observations. 
However, this does not imply that we have in advance specified 
the equation of state of dark energy. In fact, $\rho_{r_c}$ in Eq.(\ref{friedmann:dgp}) 
is not unknown function: the scale-factor dependence is known and $r_c$ is written in terms of 
$\Omega_{K0}$ and $\Omega_{M0}$. Hence in DGP the equation of state of dark energy 
is already specified.}
\beqa
1+{1\over 3\beta}={4\Omega_M(a)^2-4(1-\Omega_K(a))^2+
2\sqrt{1-\Omega_K(a)}(3-4\Omega_K(a)+2\Omega_M(a)\Omega_K(a)+\Omega_K(a)^2)\over
3\Omega_M(a)^2-3(1-\Omega_K(a))^2+
2\sqrt{1-\Omega_K(a)}(3-4\Omega_K(a)+2\Omega_M(a)\Omega_K(a)+\Omega_K(a)^2)},
\label{beta}
\eeqa
where 
\beqa
\Omega_M(a)={8\pi G\rho_M\over 3H(a)^2}
&&={\rho_M\over \left(\sqrt{\rho_{r_c}}+\sqrt{\rho_{r_c}+\rho_M}\right)^2-
{3K\over 8\pi Ga^2}}\nonumber\\
&&={\Omega_{M0}(a/a_0)^{-3}\over \left({1-\Omega_{M0}-\Omega_{K0}\over 2\sqrt{1-\Omega_{K0}}}
+\sqrt{{(1-\Omega_{M0}-\Omega_{K0})^2\over 4(1-\Omega_{K0})}
+\Omega_{M0}(a/a_0)^{-3}}\right)^2 +\Omega_{K0}(a/a_0)^{-2}},
\label{omega(a)}
\eeqa
where we have used 
$\Omega_{c}=8\pi \rho_{r_c}/3H_0^2=(1-\Omega_{M0}-\Omega_{K0})^2/4(1-\Omega_{K0})$. 
$\Omega_K(a)=-K/a^2H(a)^2$. 
Thus the evolution of density perturbations is determined once we specify two parameters: 
$\Omega_{M0}$ and $\Omega_{K0}$. 
Henceforth, for simplicity of presentation, we restrict 
ourselves to a flat universe. But the analysis is easily extended to non-flat universes 
straightforwardly. In a flat universe Eq.(\ref{beta}) is further simplified
\beqa
1+{1\over 3\beta}={2+4\Omega_M(a)^2\over 3(1+\Omega_M(a)^2)}.
\eeqa

As in the case of general relativity, we rewrite Eq.(\ref{linear:modified}) 
by changing the argument from the cosmic time to the scale factor $a$: 
\beqa
{dH(a)^2\over da}+2\left({3\over a}+{d^2\delta/da^2\over d\delta/da}\right)H(a)^2
={2\Omega_{M0}H_0^2a_0^3\delta\over a^5 (d\delta/da)}
\left({1+2\Omega_M(a)^2\over 1+\Omega_M(a)^2}\right).
\eeqa
Putting $\Omega_M(a)=1$ (or $\Omega_M(a)+\Omega_K(a)=1$ for non-flat universes) 
recovers the equation in general relativity. 
Quite similar to the case of general relativity, 
we solve the above equation for $H(a)^2$ to obtain
\beqa
\left({H\over H_0}\right)^2=
2\Omega_{M0}{(1+z)^2\over \delta'(z)^2}\int_z^{\infty}{\delta\over 1+z}
(-\delta') \left({1+2\Omega_M(z)^2\over 1+\Omega_M(z)^2}\right)dz.
\eeqa
Putting $z=0$ gives an implicit equation for $\Omega_{M0}$ (or $\Omega_{M0}$ and 
$\Omega_{K0}$ for non-flat universes)
\beqa
1={2\Omega_{M0}\over \delta'(0)^2}\int_0^{\infty}{\delta\over 1+z}
(-\delta') \left({1+2\Omega_M(z)^2\over 1+\Omega_M(z)^2}\right)dz.
\eeqa
Then we can rewrite the solution 
\beqa
\left({H\over H_0}\right)^2=
{(1+z)^2\delta'(0)^2\over \delta'(z)^2}
-2\Omega_{M0}{(1+z)^2\over \delta'(z)^2}\int^z_{0}{\delta\over 1+z}
(-\delta')\left({1+2\Omega_M(z)^2\over 1+\Omega_M(z)^2}\right)dz,
\label{hubble:linear:dgp}
\eeqa
which is the DGP counterpart of Eq.(\ref{hubble:linear2}). Here $\Omega_M(z)$ 
is  defined by Eq.(\ref{omega(a)}). 

Equating Eq.(\ref{hubble:linear:dgp}) with Eq.(\ref{hubble:dl}) gives the 
consistency relation between observables in DGP in a flat universe:
\beqa
{1\over H_0^2(dr(z)/dz)^2}=
{(1+z)^2\delta'(0)^2\over \delta'(z)^2}
-2\Omega_{M0}{(1+z)^2\over \delta'(z)^2}\int^z_{0}{\delta\over 1+z}
(-\delta')\left({1+2\Omega_M(z)^2\over 1+\Omega_M(z)^2}\right)dz,
\eeqa
which should hold if DGP model is the correct theory of gravity 
in the universe. \footnote{Again in non-flat universes, the CMB shift parameter may be used 
to break the degeneracy with $\Omega_{K0}$.}  

\subsection{Reconstructing $H(z)$ from $\delta$: DGP vs. GR}

In order to demonstrate how the breakdown of the consistency relation Eq.(\ref{consistency}) 
occurs, we calculate $H(z)$ from $\delta$ using Eq.(\ref{hubble:linear2}) when 
the correct theory of gravity is not general relativity and compare it with the modified 
Friedmann equation Eq.(\ref{friedmann:dgp}) which can be determined by the distance 
measurements Eq.(\ref{hubble:dl}). In short, we compare the right hand side and the left
hand side of Eq.(\ref{consistency}). More concretely, we consider the case when the true 
cosmology is a flat DGP model Eq.(\ref{friedmann:dgp}) and prepare $\delta$ (data) 
for $\Omega_{M}=0.3,0.2$ using Eq.(\ref{linear:modified}).  But we wrongly assume the true 
cosmology to be a flat FRW model in GR and determine $\Omega_{M0}$ using Eq.(\ref{omega}) and 
calculate $H(z)$ from Eq.(\ref{hubble:linear2}). We do not include the effects of 
observational uncertainties which will be considered in subsequent work. 

\begin{figure}
\includegraphics[width=13cm]{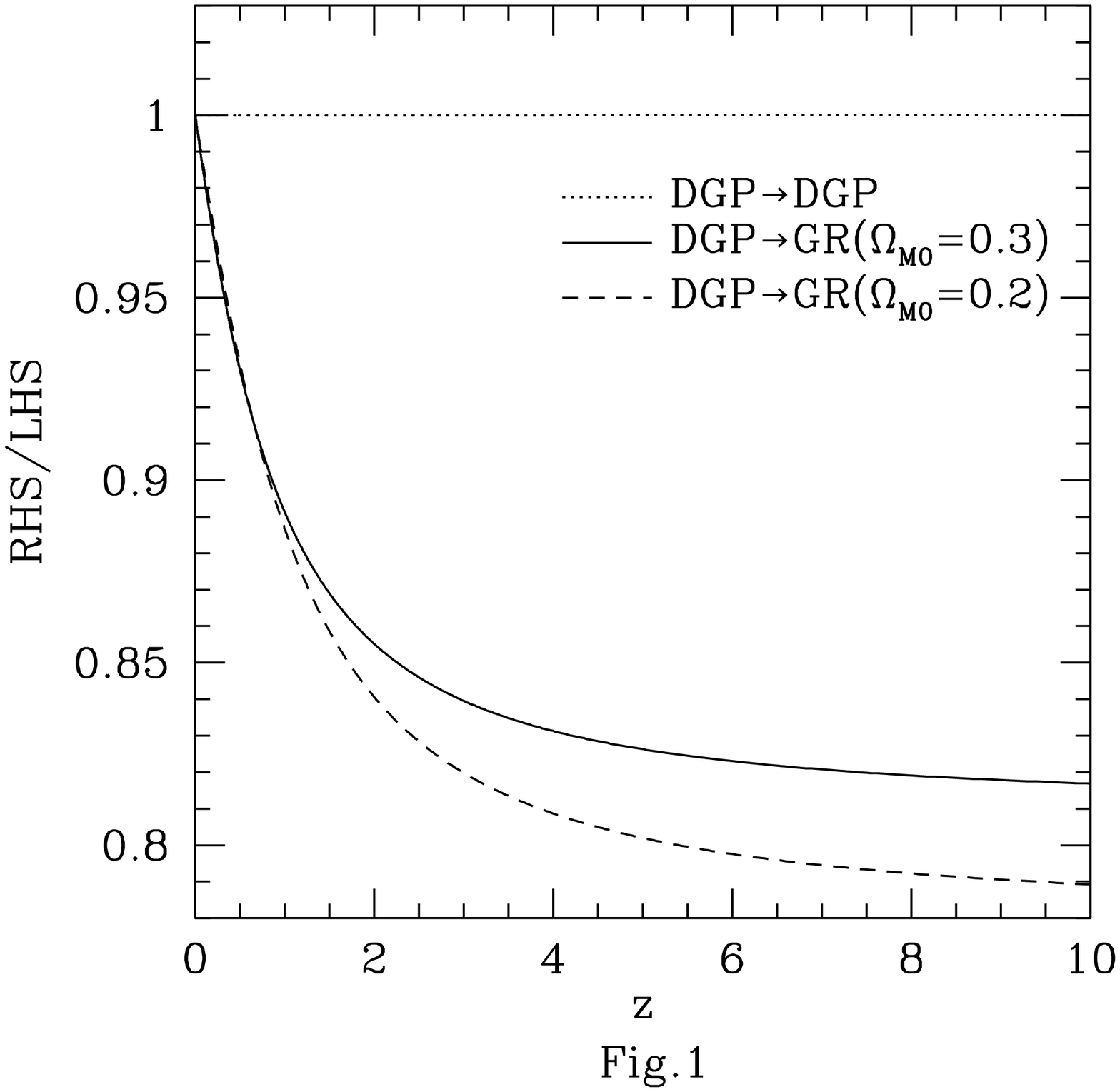} 
\caption{The ratio of RHS to LHS of Eq.(\ref{consistency}). We assume true cosmology 
to be a flat DGP model and reconstruct $H(z)$ using general 
relativity Eq.(\ref{hubble:linear2}). The solid (dashed) curve is for $\Omega_{M0}=0.3 (0.2)$. 
 The dotted line is the result of the case when $H(z)$ is reconstructed  using 
DGP Eq.(\ref{hubble:linear:dgp}). }
\label{fig1}
\end{figure}

The results are shown in Fig. \ref{fig1}. $H(z)^2$ calculated from $\delta$ 
using Eq.(\ref{hubble:linear2}) (RHS) is compared with Eq.(\ref{friedmann:dgp}) (LHS). 
The solid (dashed) curve is for $\Omega_{M0}=0.3 (0.2)$ showing that about 20\% 
differences are expected. The curve deviates from unity for $z\siml 5$ and 
the difference is saturated beyond that since the universe is matted dominated then. 
The dotted line is the result of the case when $H(z)$ is determined by 
using the correct equation Eq.(\ref{hubble:linear:dgp}).  
$\Omega_{M0}$ determined from Eq.(\ref{omega}) is $0.243 (0.156)$ for 
$\Omega_{M0}=0.3 (0.2)$, 
respectively, which also differs by about 20\%. It gives another consistency test of 
the cosmological model. 

According to the analysis of the simulated future weak lensing data (like LSST\cite{lsst}) 
in \cite{knox}, the distances would be measured within 1\% out to $z\simeq 3$ (the error is 
similar for SNAP \cite{snap} but out to $z\simeq 2$) and  
the growth rates would be determined within 4\% at $z\leq 1.2$. 
So, we expect the left hand side of Eq.(\ref{consistency}) would be determined within 
$\sim 2\times 1\%$, while the right hand side within $\sim 4\times 4=16\%$ at $z\leq 1.2$. 
The detailed analysis is left as our future work. 

\section{Summary}

In this paper, we have derived a consistency relation in general relativity which relates 
the distance and the density perturbation assuming perturbative properties of dark energy. 
Breaking of the consistency relation would be the signature of the breakdown of the assumptions: 
general relativity is not the correct theory of gravity at the cosmological scale or 
dark energy has unusual properties (unusual sound speed or anisotropic stress or interaction 
with dark matter). 
 
Four cosmological observables (the amplitudes and the spectral indices of scalar/tensor 
perturbations) out of three inflationary parameters (the energy scale, two slow-roll parameters 
$\epsilon$ and $\eta$) gives a consistency relation for single-field inflation \cite{kolb}. 
It represents an extremely distinctive signature of inflation. 
The verification of the relation would be the direct proof of (single-field) inflation and 
would be the milestone of inflationary cosmology, although it would be difficult to 
verify the relation in the foreseeable future. 

Likewise, the proof or disproof of the consistency relation in cosmology would give us 
a clue as to the nature of dark energy or the nature of gravity on cosmological scales which 
would not been obtained by local experiments. As such, any observational methods to test it 
should be welcome. Twentieth century clouds \cite{jim2} should have a silver lining. 
The consistency relation would help to clear up the dark clouds of the cosmos.

\begin{acknowledgments}
TC would like to thank RESCEU at the University of Tokyo for hospitality. 
This work was supported in part by a Grant-in-Aid for Scientific 
Research (No.17204018) from the Japan Society for the 
Promotion of Science and in part by Nihon University. 
\end{acknowledgments}


\end{document}